\documentclass[a4paper,amsmath,amssymb,prl,floatfix,footinbib,twocolumn,preprintnumbers]{revtex4-1}

\usepackage[dvips]{graphicx}
\usepackage{dcolumn}
\usepackage{bm}
\usepackage{dsfont}
\usepackage{color}
\usepackage[normalem]{ulem} 
\usepackage{url}
\usepackage[colorlinks=true ,citecolor=blue]{hyperref}
\usepackage{amsmath,amssymb,bbold}

\def\b{\mathbf}
\def\rm{\mathrm}
\def\pa{\partial}

\newcommand{\nn}{\nonumber}
\newcommand{\be}{\begin{equation}}
\newcommand{\ee}{\end{equation}}
\newcommand{\bea}{\begin{eqnarray}}
\newcommand{\eea}{\end{eqnarray}}
\newcommand{\mv}[1]{\langle #1\rangle}
\newcommand{\f}{\frac}
\newcommand{\bra}{\langle}
\newcommand{\ket}{\rangle}

\begin{document}

\title{Topological Varma superfluid in optical lattices}

\author{M. Di Liberto$^{1,2}$, A. Hemmerich$^{3,4,5}$, and C. Morais Smith$^{1,5}$}
\affiliation{$^1$ Institute for Theoretical Physics, Centre for Extreme Matter and Emergent Phenomena, Utrecht University, Leuvenlaan 4, 3584CE Utrecht, the Netherlands \\
$^2$INO-CNR BEC Center and Dipartimento di Fisica, Universit\`a di Trento, 38123 Povo, Italy\\
$^3$Institut f\"{u}r Laser-Physik, Universit\"{a}t Hamburg, LuruperChaussee 149 22761 Hamburg, Germany \\
$^4$The Hamburg Centre for Ultrafast Imaging, Luruper Chaussee 149, 22761 Hamburg, Germany\\
$^5$Wilczek Quantum Center, Zhejiang University of Technology, Hangzhou, China}



\begin{abstract}
Topological states of matter are peculiar quantum phases showing different edge and bulk transport properties connected by the bulk-boundary correspondence. While non-interacting fermionic topological insulators are well established by now and have been classified according to a ten-fold scheme, the possible realisation of topological states for bosons has not been much explored yet. Furthermore, the role of interactions is far from being understood. Here, we show that a topological state of matter exclusively driven by interactions may occur in the p-band of a Lieb optical lattice filled with ultracold bosons. The single-particle spectrum of the system displays a remarkable parabolic band-touching point, with both bands exhibiting non-negative curvature. Although the system is neither topological at the single-particle level, nor for the interacting ground state, on-site interactions induce an anomalous Hall effect for the excitations, carrying a non-zero Chern number. Our work introduces an experimentally realistic strategy for the formation of interaction-driven topological states of bosons.
\end{abstract} 

\maketitle

\emph{Introduction.} The seminal work of Kane and Mele for graphene in the presence of a strong spin-orbit coupling \cite{Kane2005} has opened the field of topological insulators and brought us a deeper comprehension of phenomena like the quantum Hall effect, which had been known for decades \cite{vonklitzing1980}. A universal classification scheme based on symmetries and dimensionality has enabled a solid basis for our understanding of non-interacting fermionic topological systems \cite{Altland1997,Ryu2010,Kane2010,Niu2010}. In contrast, interaction-driven topological states of matter have remained elusive. Topological phases were originally predicted to occur in lattice models with dominating next-nearest-neighbor interaction \cite{Zhang2008}, a scenario which appears difficult to envisage in real experiments. More precise exact diagonalization \cite{Daghofer2014} and DMRG \cite{Motruk2015} calculations, however, contradicted these mean-field results. 

Predictions of topological phases in Dirac-like materials exhibiting a linear dispersion suffer from the additional drawback that theses systems are genuinely quantum critical: due to the zero density of states at the neutrality point, only perturbations exceeding a certain critical value are able to induce a topological phase. On the other hand, systems with a parabolic band touching point may become topological for infinitesimal values of the interaction \cite{Sun2009}. A paradigmatic example is Bernal-stacked bilayer graphene \cite{Falko2006}, but other cases have also been identified, and this field has attracted much interest recently \cite{Sun2012, Ortix2015}. 

The search for {\it interaction-driven} topological systems has mostly concentrated on electronic condensed matter \cite{Zhang2008, Sun2009, Daghofer2014, Motruk2015, MoraisSmith2015, Ortix2015} or on the equivalent cold-atom fermion system \cite{Sun2012,Dauphin2012,Dauphin2015}. Only recently, the corresponding bosonic analogs became the subject of theoretical investigations  \cite{Lind2011,Wen2012,Senthil2013,Liu2016,Zhu2016}. As for experiments, considerable progress has been achieved in realizing topological states with cold atoms (fermions or bosons) in the {\it non-interacting} regime. Examples range from the Su-Schrieffer-Heeger model \cite{Atala2013}, to the Harper-Hofstadter model \cite{Aidel2013, ketterle2013} or the long-sought Haldane model \cite{esslinger2014}. In these systems, topological features of the bands, like a non-trivial Zak phase, Berry curvature, or Chern number, characterize the non-interacting model and interactions play only a marginal role \cite{Brandes2015,Gremaud2015,LeHur2015}. In this context the intriguing question arises, how to engineer a parabolic band touching with cold atoms, because in this case topological features might emerge exclusively due to interactions. The main difficulty is that both parabolic bands must have a non-negative curvature in order to provide a band minimum, where the bosons can be condensed.

Here, we show that this scenario can be achieved by employing interacting ultracold bosons in a two-dimensional Lieb lattice, known as a model of the copper oxide planes of cuprate superconductors. The desired band touching arises between the second and third bands at the $M$-point of the Brillouin zone.  We show that with proven experimental techniques a stable Bose-Einstein condensate (BEC) can be formed at this high-symmetry point, which exhibits macroscopic angular momentum and superfluid plaquette currents, such that time-reversal symmetry is broken, but translational symmetry is preserved. This phase, which turns out to be topologically trivial, is the bosonic analog of a phase first proposed by Varma to describe the pseudogap regime of high-$T_c$ cuprates \cite{Varma1997,Varma1999,Varma2006,Varma2012}. Surprisingly, when we calculate the band structure of the Bogolyubov excitations, we find that they exhibit emergent topological properties that were absent in the non-interacting system. 

\begin{figure}[!tbp]
\includegraphics[width=1.0\columnwidth]{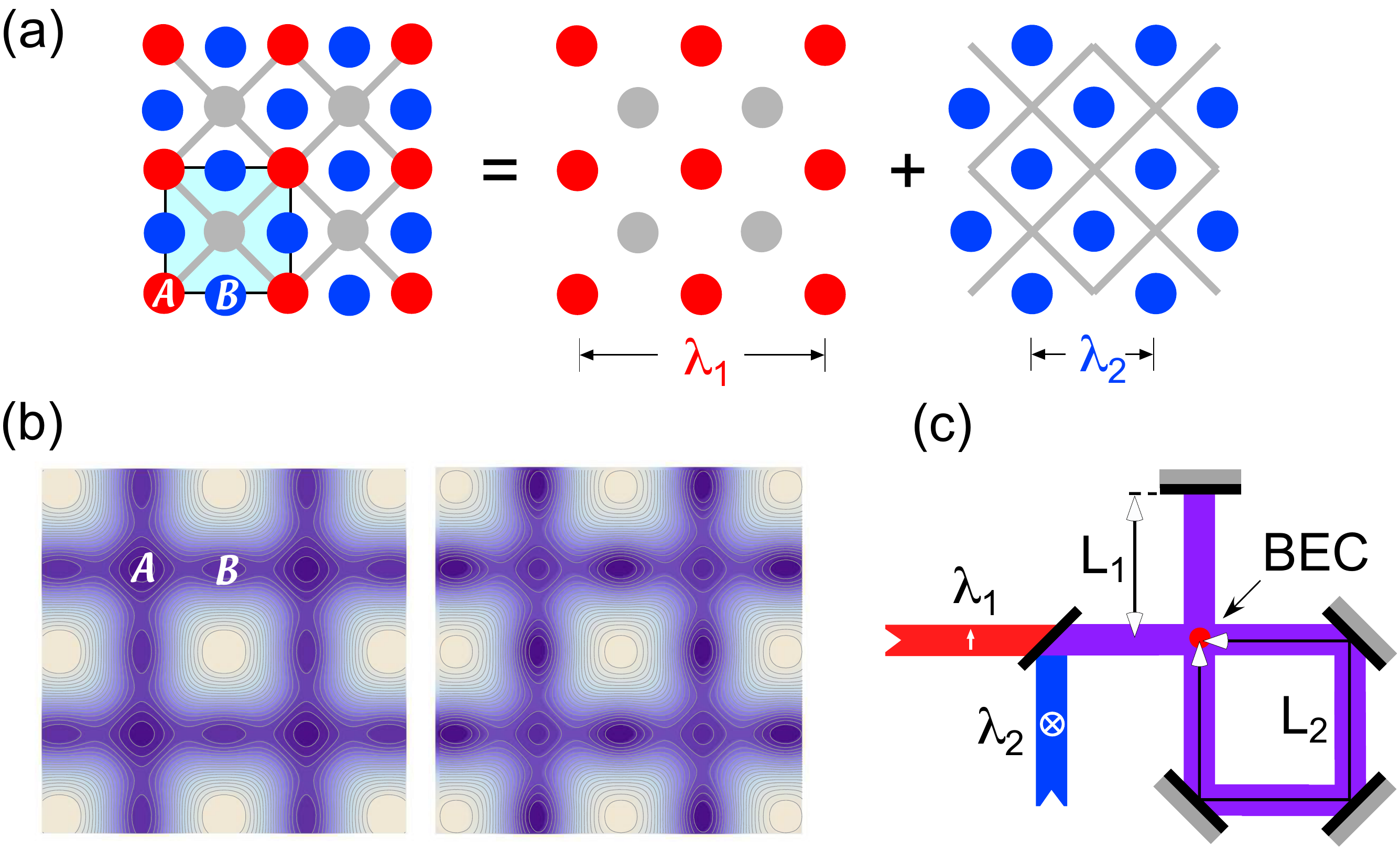}
\caption{(a) A face-centered square lattice (left detail) is obtained by superimposing the two square lattices shown in the center and in the right detail (with $\lambda_1 \approx 2\lambda_2$). Colored disks show potential minima, gray disks or lines show maxima. (b) By varying the relative depth of the superimposed square lattices, the relative depth of the $\mathcal{A}$ and the $\mathcal{B}$ wells of the resulting Lieb lattice can be tuned. (c) Experimental realization of a Lieb-lattice potential with built-in band swapping functionality (see text for details).}
\label{fig_lattice}
\end{figure}

{\emph{Optical lattice.}} In contrast to previous realizations of a Lieb lattice with light-shift potentials \cite{Tak:15}, our implementation includes the necessary machinery to prepare atoms in the second Bloch band, and hence to include orbital degrees of freedom. The Lieb-lattice geometry, depicted in Fig.~\ref{fig_lattice}(a), arises if a unit cell (light blue rectangle in the left detail) with two classes of sites denoted $\mathcal{A}$ and $\mathcal{B}$ is translated via a square Bravais lattice. In order to excite atoms into the second band of this lattice using the technique demonstrated in Refs.~\cite{Wir:11,Oel:13}, additional functionality (henceforth referred to as band swapping) is required, which allows one to rapidly switch the well depths of the corner sites $\mathcal{A}$ and the bond sites $\mathcal{B}$. Experimentally, this may be achieved by superimposing the two potentials
\bea \label{V}
V_1(x,y) &\equiv& -V_{1,0} [\cos^2(k_1 x)+\cos^2(k_1 y)] \,,\\
V_2(x,y) &\equiv& -\frac{V_{2,0}}{4} |\cos(k_2 x)+\cos(k_2 y)|^2\,, \nn
\eea
thus obtaining $V(x,y) \equiv V_1(x,y) + V_2(x,y)$. Here, $k_{\nu} \equiv 2\pi/\lambda_{\nu}$ with $\lambda_{\nu}, \nu \in \{1,2\}$ denoting the wavelengths of the light fields forming $V_{\nu}(x,y)$ and $\lambda_{1} \approx 2 \lambda_{2}$. As illustrated in the central and right details of Fig.~\ref{fig_lattice}(a), both potentials $V_1(x,y)$ and $V_2(x,y)$ represent simple square lattices, however, rotated with respect to each other by $45^{\circ}$, and with lattice constants differing by a factor $\sqrt{2}$. Indicating their potential  minima by red and blue disks and their potential maxima by gray disks or grey lines, respectively, Fig.~\ref{fig_lattice}(a) immediately clarifies how their superposition yields the desired Lieb-lattice geometry. By tuning the ratio $V_{1,0}/V_{2,0}$ around unity, the $\mathcal{A}$ wells can be tuned to be deeper than the $\mathcal{B}$ wells or vice versa. This is shown in Fig.~\ref{fig_lattice}(b), where the potential is plotted for $V_{1,0}/V_{2,0} = 0.85$ and $V_{1,0}/V_{2,0} = 1.17$ on the left- and right-hand sides, respectively.

Experimentally, the required lattice is produced by the setup shown in Fig.~\ref{fig_lattice}(c). Two laser beams with wavelengths $\lambda_1$ and $\lambda_2 = \frac{1}{2}\lambda_1 + \delta\lambda$, both detuned to the negative side of an atomic transition with $\delta\lambda$ corresponding to a few ten MHz, are retro-reflected from the same mirror. The polarization of the beam at wavelength $\lambda_1$ is adjusted to be linear within the lattice plane. Hence, no interference occurs at the crossing point, where an atomic BEC is prepared, such that the potential $V_1(x,y)$ arises. Similarly, the beam at wavelength $\lambda_2$ exhibits linear polarization perpendicular to the lattice plane, such that maximal interference at the crossing point yields the potential $V_2(x,y)$. The two potentials can be precisely positioned relative to each other (to better than a few nm) by adjusting $\delta\lambda$ and the ratio between the distances $L_1$ and $L_2$ indicated in the Fig.~\ref{fig_lattice}(c) \cite{Elm:05}.\\

\begin{figure}[t]
\includegraphics[width=1.0\columnwidth]{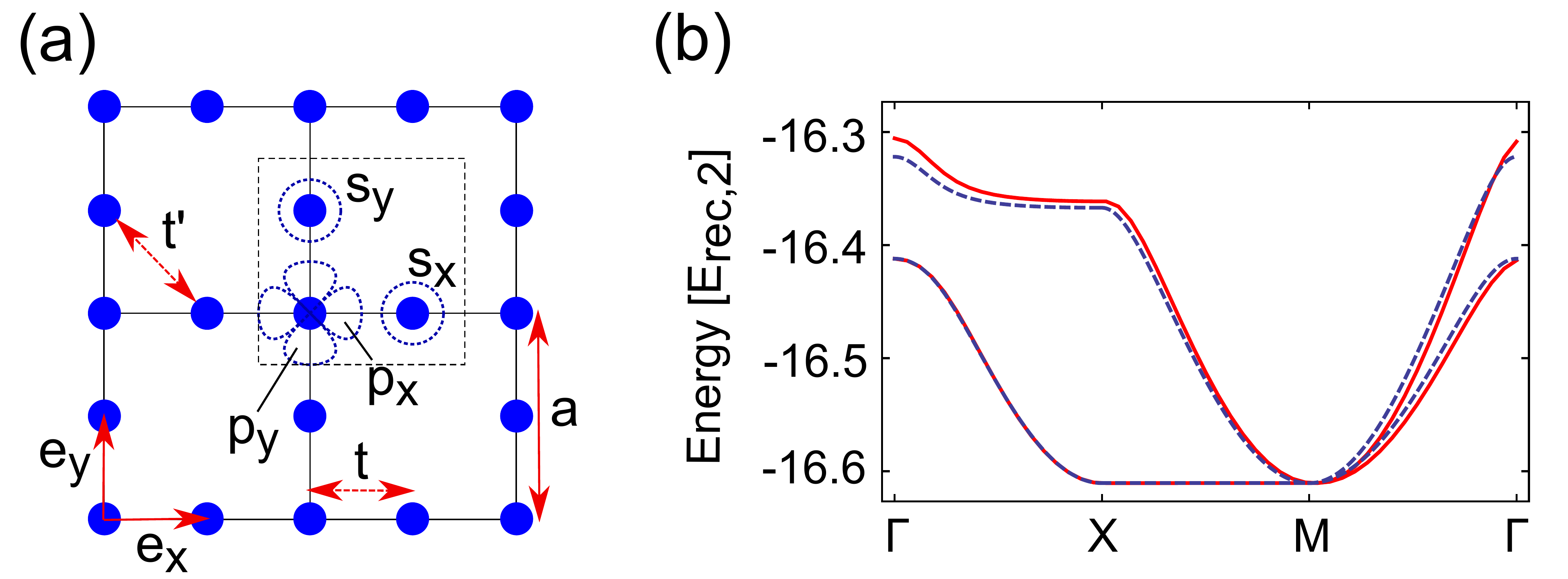}
\caption{(a) Tight-binding lattice model. (b) Comparison between the tight-binding band structure (dashed blue line) and the exact band structure (continuous red line) for $V_{1,0} = 10.6\, E_\textrm{rec,2} $ and $V_{2,0} = 9.4\, E_\textrm{rec,2} $. The fitting parameters $t = 0.166$, $t' = 0.011$, $\xi_p = 0.209$ and overall energy shift $E_0 = -16.367$ are given in units of the recoil energy $E_\textrm{rec,2} \equiv \hbar^2 k_2^2/(2m)$, with $m$ the atomic mass.}
\label{fig_TBmodel}
\end{figure}

\emph{Theoretical model} We consider the case when the corner sites ($\mathcal{A}$) are significantly deeper than the bond sites ($\mathcal{B}$) such that the former host $p_x$ and $p_y$ orbitals, while the latter host $s$ orbitals. The low-lying $s$ orbitals of the corner sites are neglected. The corresponding minimal tight-binding Hamiltonian reads
\begin{align} \label{H0}
H^0 =&\,-t \sum_{\b{r},\alpha,\sigma} \left( \sigma \, p^\dag_{\alpha,\b{r}} s_{\b{r} + \sigma \b{e}_\alpha} + \rm{h.c.}\right)\\
& - t'  \sum_{\b{r},\rho,\sigma} \left( s^\dag_{\b{r} + \sigma \b{e}_x} s_{\b{r} + \rho \b{e}_y} + \rm{h.c.}\right) \nn\\
& + \xi_p \sum_{\b{r},\alpha}  p^\dag_{\alpha,\b{r}}  p_{\alpha,\b{r}} + \xi_s \sum_{\b{r},\alpha} s^\dag_{\b{r} + \b{e}_\alpha} s_{\b{r} +  \b{e}_\alpha}\nn\,,
\end{align}
where $t$ is the nearest-neighbor hopping between $s$ and $p$ orbitals, $t'$ is the next-nearest-neighbor hopping between $s$ orbitals, and $\xi_p$ and $\xi_s$ are, respectively, the on-site energies for $p$ and $s$ orbitals (see Fig.~\ref{fig_TBmodel}(a)). We will consider $\xi_s = 0$ in the remainder of this work. Let us define $s_{\alpha,\b{r}}\equiv s_{\b{r}+\b{e}_\alpha}$ and introduce the vector $\Psi_{\b k} \equiv (p_{x,\b{k}},p_{y,\b{k}},s_{x,\b{k}},s_{y,\b{k}})$. In momentum space, the Hamiltonian can be written as $H^0 = \sum_{\b{k}} \Psi^\dag_{\b{k}} H^0_{\b{k}} \Psi_{\b{k}}$, where
\be
\label{ham4b}
H^0_{\b{k}}=
\begin{pmatrix}
\xi_p & 0 & -2it \tilde s_x & 0 \\
0 & \xi_p & 0 & -2it \tilde s_y & \\
2it \tilde s_x & 0 & \xi_s & -4 t' \tilde c_x \tilde c_y\\
0 & 2it \tilde s_y & -4 t' \tilde c_x \tilde c_y & \xi_s
\end{pmatrix}\,.
\ee
Here, we have defined $\tilde c_{x,y} \equiv \cos(k_{x,y}/2)$, $\tilde s_{x,y} \equiv \sin(k_{x,y}/2)$, and introduced units for which the lattice constant $a=1$. At the high-symmetry points, the lowest-energy eigenvectors have the form (up to a normalisation constant): $\Psi_X = \left( - i \epsilon,0,2t,0 \right)$, $\Psi_{X'} = \left( 0, -i \epsilon,0,2t \right)$, $\Psi_{M}^{(1)} = \Psi_X$, $\Psi_{M}^{(2)} = \Psi_{X'}$, corresponding to the energy $\epsilon = \left( \xi_p - \sqrt{16t^2+\xi_p^2} \right)/2$. The $M$ point is two-fold degenerate.

A comparison between the band structure derived from the Hamiltonian in Eq.~(\ref{H0}) and from the experimental potential given in Eq.~(\ref{V}) is shown in Fig.~\ref{fig_TBmodel}(b). The desired quadratic-band touching at the high-symmetry point $M$ of the Brillouin zone, arising from two bands with non-negative curvature, may be promptly observed upon inspection of the figure. 

In the presence of contact interactions $U_s$ and $U_p$ for the $s$- and $p$-orbitals, respectively, and within the harmonic approximation, the interaction Hamiltonian can be cast in the form \cite{Bloch2008,Liu2006}
\be
\label{int}
H_\rm{int} = \f{U_s}{2} \sum_{\b r,\alpha} n_{s_\alpha,\b r} (n_{s_\alpha,\b r} - 1 ) + \f{U_p}{2} \sum_{\b r} \left(n_{p,\b r}^2 - \f{L_{z,\b r}^2}{3} \right)\,,\nn
\ee
where $n_{s_\alpha,\b r} = s^\dag_{\alpha,\b{r}}s_{\alpha,\b{r}}$, $n_{p,\b r} \equiv n_{p_x,\b{r}} + n_{p_y,\b{r}}$ and the angular-momentum operator is given by $L_{z,\b r} = -i \left(p^\dag_{x,\b{r}} p_{y,\b{r}}-p^\dag_{y,\b{r}}p_{x,\b{r}} \right)$. Inspection of the Hamiltonian shows that repulsive interactions favour a ground state with finite angular momentum $\mv{L_{z,\b r}} \neq 0$, thus breaking time-reversal symmetry $\cal T$. \\


\begin{figure}[t]
\includegraphics[width=0.95\columnwidth]{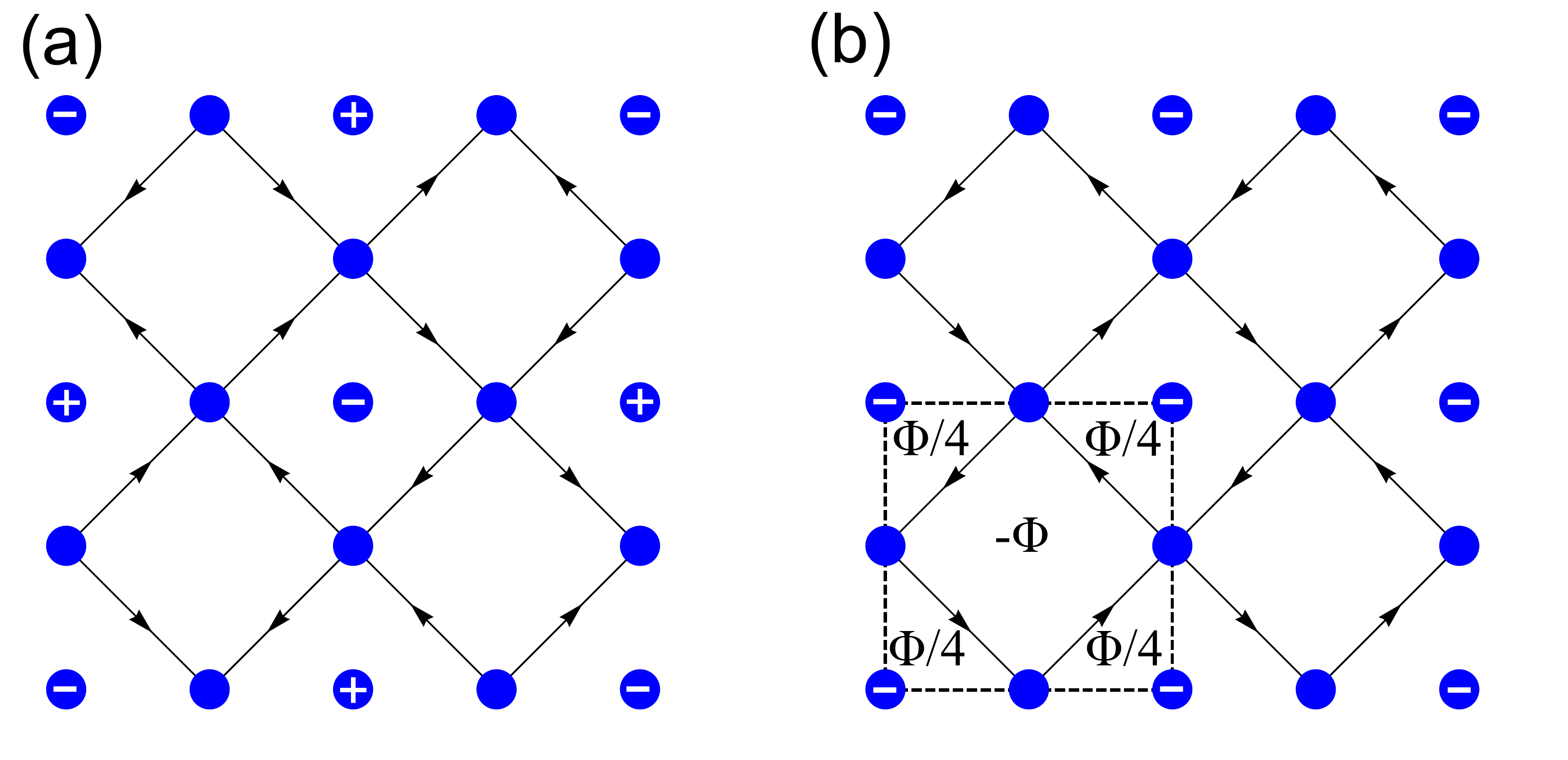}
\caption{(a) $X-X'$ condensate: a pattern of staggered angular momenta and currents arises (b) $M$ condensate: angular momenta are rectified and currents form loops in the elementary plaquettes (dashed line).}
\label{fig_groundstate}
\end{figure}

{\emph{Ground-state.}} Lattice systems exhibiting a degenerate set of minima favour condensation at the high-symmetry points, as for instance found for the Kagome lattice \cite{Zhai2012}. Therefore, we distinguish two types of condensates, depending on whether the condensation occurs at the points $X-X'$ or at the two-fold degenerate point $M$. This scenario is confirmed by solving the Gross-Pitaevskii equation in imaginary time, which in addition allows us to exclude simultaneous condensation at all the three points. At the momenta corresponding to the high-symmetry points $X$, $X'$, and $M$, only some orbitals have a non-vanishing amplitude for the Hamiltonian in Eq.~(\ref{H0}). These considerations lead us to propose the following ansatz for the condensate wave-function
\be
\mv{p_{\alpha,\b{r}}} = \sqrt{\rho_p} e^{i\theta_\alpha} e^{i\b{k}_{\alpha}\cdot \b{r}} \,,\quad \mv{s_{\alpha,\b{r}}} = \sqrt{\rho_s} e^{i\phi_\alpha} e^{i\b{k}_\alpha\b{\cdot (r}+\b{e}_\alpha)} \,.\nn
\ee
For the $X-X'$ condensate, one has $\b k_x = X = (\pi,0)$ and $\b k_y = X' = (0,\pi)$. For the $M$ condensate, $\b k_x = \b k_y = M = (\pi,\pi)$. We make a gauge choice: $\phi_y=0$. Minimisation of the mean-field free energy determines the phases $\theta_x = \pi$, $\theta_y = \pi/2$, $\phi_x = \pi/2$, $\phi_y = 0$ and the mean-field free energy reads 
\be
E_{\rm{MF}} = -8t\sqrt{\rho_p \rho_s} + 4U_p\rho_p^2/3 + U_s\rho_s^2 + 2\xi_p\rho_p\,.\nn
\ee
Since the condensates break time-reversal symmetry $\cal T$, each solution is two-fold degenerate. The time-reversal conjugate solution can be obtained by substituting $\phi_x \rightarrow - \phi_x$ and $\theta_y \rightarrow -\theta_y$.

The angular momentum expectation value is $\mv{L_{z,\b r}} = -2\rho_p(-1)^{m+n}$ in the $X-X'$ phase and $\mv{L_{z,\b r}} = -2\rho_p$ in the $M$ phase, where $m$ and $n$ are integers corresponding to $\b r = (m,n)$. Hence, the overall angular momentum vanishes for the  $X-X'$ phase but not for the $M$ phase. The $X-X'$ condensate breaks the translational symmetry of the Bravais lattice, whereas the $M$ condensate does not (see Fig.~\ref{fig_groundstate}). Phase gradients induce superfluid currents. The bond-current operator is defined as $J^{\mu\nu}_{ij}=-i t_{ij} \left( b^\dag_{\mu,i} b_{\nu,j} - b^\dag_{\nu,j} b_{\mu,i} \right),$ where $b^\dag_{\mu,i}$ is the creation operator for a boson at site $i$ and flavor/orbital $\mu$ and $t_{ij}$ is the hopping amplitude between sites $i$ and $j$. There are currents only along the $s-s$ bonds with amplitude $|J^{ss}| = 2 t' \rho_s$. 

The ground states may possess non-trivial topological properties, such as a Hall response. The calculation of the transverse (Hall) conductivity $\sigma_{xy}$ revealed that this is not the case (see Refs. \cite{Avron1985,Lind2011,suppmat} for more details). However, as we shall discuss below, topological properties arise for the excitations of the $M$ condensate.

Corrections to the ground-state mean-field energy given by the zero-point fluctuations of the Bogolyubov modes have been also investigated following the procedure outlined in Ref.~\cite{Zhai2012} and we found that the $M$ condensate has a larger energy compared to the $X-X'$ condensate, of the order of 1 pK \cite{suppmat}. This small value makes us confident that a metastable $M$ condensate could be experimentally realizable. Nevertheless, the precise estimation of the condensate lifetime is a highly non-trivial problem, which we defer to future investigations. 

\begin{figure}[t]	
\begin{center}
\includegraphics[width=1.0\columnwidth]{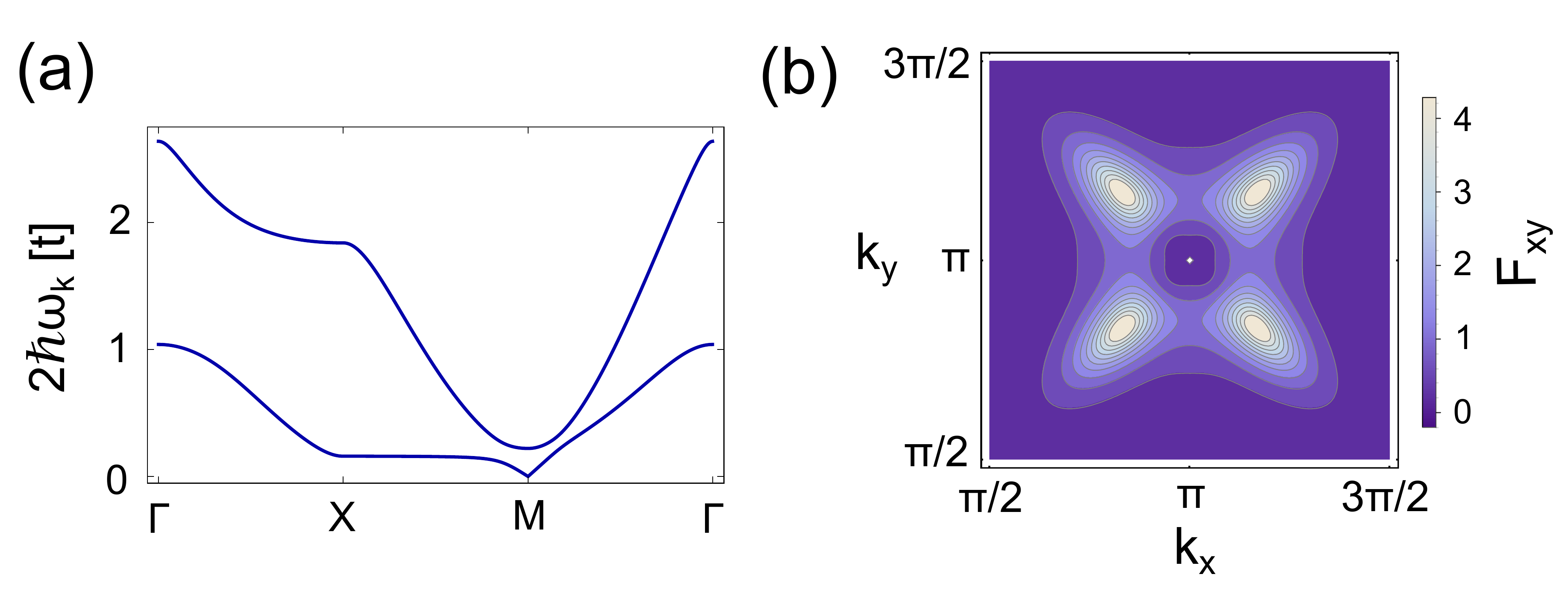}
\end{center}
\vspace{-.5cm}
\caption{(a) Bogolyubov spectrum. Parameters are chosen in units where $t=1$: $t'=0.2$, $\xi_p = 1.5$, $U=0.005$. Numerical minimization of the mean-field free energy under the constraint $\rho_s + \rho_p = 100$ yields $\mu = -1.01882$, $\rho_p U = 0.1730$ and $\rho_s U = 0.3269$. (b) Berry curvature $F_{xy}$ of the lowest branch of the Bogolyubov spectrum. The white spot at $M = (\pi, \pi)$ indicates that the Berry curvature is not calculated at this point, since this is the momentum at which condensation occurs.}
\label{BogMp}
 \end{figure}


{\emph{Excitations and topology.}} The quadratic band-touching appearing at the $M$-point in the non-interacting spectrum is protected by inversion and time-reversal symmetry, and it exhibits a $2\pi$ Berry flux \cite{Sun2009,Sun2012,suppmat}. Upon condensation, the band degeneracy is lifted  because the U(1) symmetry breaking requires the existence of only one gapless mode, according to Goldstone's theorem. Therefore, the band splitting appearing for the excitations and the further breaking of time-reversal symmetry can allow for a non-vanishing Berry curvature and Chern number. 

Within the Bogolyubov approximation, the Hamiltonian can be written as $H=H^0 + H^{\rm{int}} \approx E_{\rm{MF}} + H^{\rm{Bog}}$, with $H^{\rm{Bog}} = \f 1 2 \sum_{\b k} \delta \psi_{\b k}^\dag H^{\rm{Bog}}_{\b k} \delta \psi_{\b k}$, where we have introduced the Nambu spinor $\delta \psi_{\b k} = (\delta p_{x, \b k}, \delta p_{y, \b k}, \delta s_{x, \b k}, \delta s_{y, \b k}, \delta p^\dagger_{x, -\b k}, \delta p^\dagger_{y, -\b k}, \delta s^\dagger_{x, -\b k}, \delta s^\dagger_{y, -\b k})$.

The Bogolyubov Hamiltonian for the $M$ condensate reads
\be
\label{bogMp}
H^{\rm{Bog}}_{\b{k}}=
\begin{pmatrix}
H^0_{\b{k}}+H^1 & \Delta \\
\Delta^* & [H^0_{-\b{k}}+H^1]^*  \\
\end{pmatrix}\,,\nn
\ee
where $H^0_{\rm{\b k}}$ was defined in Eq.~(\ref{ham4b}), $H^1=\f 8 3 U\, \textrm{diag}(\rho_p,\rho_p,\rho_s,\rho_s) - \mu\, \mathbb 1_{4\times4}$,
\be
\Delta = 
\begin{pmatrix}
\f 2 3 U \rho_p  & -\f 2 3 i U\rho_p & 0 & 0 \\
 -\f 2 3 i U\rho_p &  -\f 2 3U \rho_p & 0 & 0 \\
0 & 0 &  \f 4 3  U\rho_s  & 0 \\
0 & 0 & 0 &  -\f 4 3  U\rho_s
\end{pmatrix}\,,\nn
\ee
and we have introduced the chemical potential $\mu$ and defined $U\equiv U_p = \f 3 4 U_s$ within the harmonic approximation (see Ref.~\cite{Liu2006}). 

The excitation spectrum is obtained by solving the eigenvalue problem $\tau_z H^{\rm{Bog}}_{\b{k}} \, W^i_{\b k} = \omega_{\b k}^i \,W^i_{\b k}$, where $\tau_z = \sigma_z \otimes \mathbb 1_{4\times4}$. The normalization of the eigenvectors is such that $W^\dagger_{\b k} \tau_z W_{\b k} = \tau_z$ and $W_{\b k} \tau_z W^\dagger_{\b k} = \tau_z$, with $W_{\b k}$ being the matrix having the eigenvectors $W^i_{\b k}$ in its columns.

\begin{figure}[t]	
\begin{center}
\includegraphics[width=1.0\columnwidth]{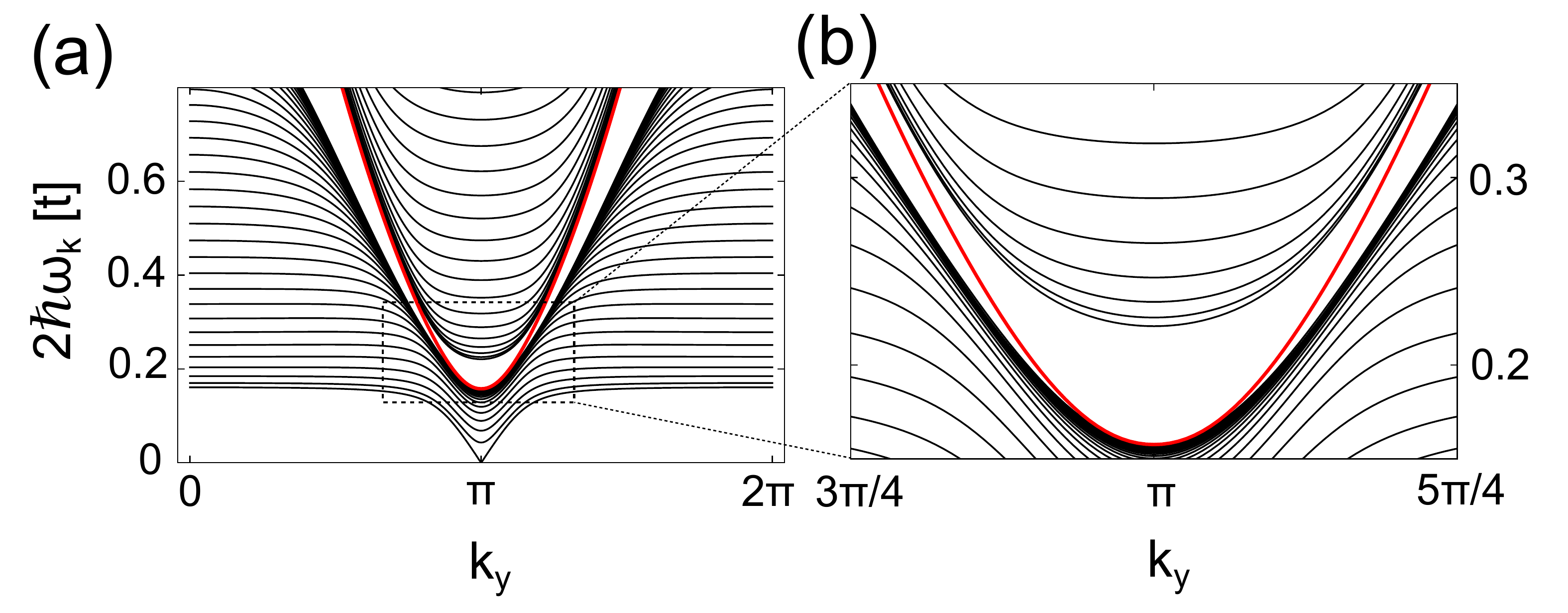}
\end{center}
\vspace{-.5cm}
\caption{(a) Bogolyubov spectrum in a cylindrical geometry with periodic boundary conditions along the $y$ direction and 36 unit cells along the $x$ direction. The red continuous line indicates the edge-state dispersion. Parameters are chosen as in Fig.~\ref{BogMp}. (b) Expanded view on the region marked by the dashed box in (a).}
\label{BogCyl}
\end{figure}

The excitation spectrum for the $M$-condensate is shown in Fig.~\ref{BogMp}(a). Notice that the lowest branch of the spectrum became non-degenerate, as a consequence of the U(1) symmetry breaking. This allows us to calculate the associated Chern number generalised for Bogolyubov Hamiltonians \cite{Shindou2013}, $c_n = \f{1}{2\pi} \int\rm d \b k \,F_{xy}(\b k)\,,$
where the Berry curvature $F_{xy}(\b k)$ reads 
$$F_{xy}(\b k) = \pa_{k_x} A_y(\b k) - \pa_{k_y} A_x(\b k)\,,\,\, A_\alpha = -i \bra W^i_{\b k} |\tau_z \pa_{k_\alpha} | W^i_{\b k} \ket\,.$$
We numerically calculated the Chern number using the procedure described in Ref.~\cite{hatsu2005} and we found $c_n = 1$ for the lowest branch of the spectrum. The Berry curvature for this branch is shown in Fig.~\ref{BogMp}(b). The non-vanishing $k$-space profile of the Berry curvature is analogous to the one of a free-particle with a quadratic band-touching point with an explicit time-reversal symmetry breaking \cite{suppmat}. The fundamental difference is that the time-reversal symmetry breaking here is {\it emergent and driven by interactions}. \\


\emph{Conclusions.} Our results provide a novel paradigm to realize topological phases of weakly-interacting bosons. By combining the condensation of atoms at a quadratic band-touching point with the spontaneous time-reversal symmetry breaking induced by the interactions, one can generate a superfluid with topological excitations. These topological excitations are non-trivial bogolons, but they are still described in terms of a single-particle Fock space \cite{fetter}. This is in contrast to purely many-body wave functions with topological properties that lie outside the Altland-Zirnbauer classification, as discussed in Ref.~\cite{kitaev2010}. The effect of interactions on the low-lying excitations is nevertheless crucial for spontaneously generating the topological properties, which are otherwise absent in the non-interacting regime.

According to the topological classification \cite{Kane2010}, this model belongs to class D: time-reversal symmetry is broken and particle-hole symmetry is intrinsically present in the Bogolyubov description. Indeed, particle-hole symmetry is represented by the operator $\mathcal C = \tau_x \mathcal K$, where $\tau_x = \sigma_x \otimes \mathbb 1_{4\times4}$ and $\mathcal K$ is the complex conjugation operator, acting as $\mathcal{C}\, \tau_z H^{\rm{Bog}}_{\b k}\, \mathcal C^\dag = - \tau_z H^{\rm{Bog}}_{-\b k}$ with $\mathcal C^2 = \mathbb 1$. A realisation of a similar class in bosonic systems has been recently proposed for a Kagome lattice of photonic crystals, where a pairing term that breaks U(1) and time-reversal symmetry is introduced by light-squeezing \cite{Clerk2015}. 

In our model, instead, interactions induce the symmetry-breaking mechanism that generates the topological features of the excitations. In particular, the non-vanishing Berry curvature characterizes an anomalous Hall effect for the excitations and affects the collective modes of the gas \cite{Spielman2012,cooper2013}. It may be directly observed with the interferometric techniques demonstrated in Refs.~\cite{cooper2012, Atala2013, esslinger2014, Aidel2015, Fla:15}, while the excitation spectra could be measured by momentum resolved Bragg spectroscopy \cite{Ern:10}. 

By diagonalizing the spectrum in a cylindrical geometry, we found the existence of excitations localised at the two edges of the cylinder, as expected from the bulk-boundary correspondence (see red line in Fig.~\ref{BogCyl}(a)-(b)). The observation of these finite-size effects could be implemented by using an optical box trap cutting the system along the symmetry axes \cite{Hadzib2013}. The absence of an overall true gap would compromise the detection of the edge states in the presence of disorder, which, however, should not be a problem in the light-shift potential of an optical lattice, since these systems are inherently clean. 

Finally, we want to recall that the ground-state of the model investigated here bears strong similarities with the Varma phase, firstly proposed for describing the pseudogap regime of high-$T_c$ superconductors \cite{Varma1997,Varma1999,Varma2006,Varma2012}. Our findings also open possibilities for new exotic states of bosons in the strongly-interacting regime \cite{Wen2012}, where an insulating Mott phase would be characterised by a gap and may also possess topological features \cite{Petrescu2013, LeHur2015}.\\

{\textbf{ Acknowledgments.}} This work was partially supported by the Netherlands Organization for Scientific Research (NWO), by the German Research Foundation DFG-SFB 925, by the Hamburg centre of ultrafast imaging (CUI), by Provincia Autonoma di Trento and by the EU-FET Proactive grant AQuS,
Project No. 640800. A. H. and C.M.S acknowledge support by NSF-PHYS-1066293 and the hospitality of the Aspen Center for Physics. We are grateful to C. Varma, J. Armaitis, L. Santos, N. Cooper, V. Juri\v{c}i\'{c}, Zi Cai, C. Ortix, H. Price and T. Ozawa for useful discussions. We especially thank N. Lindner for very insightful suggestions and a critical reading of the manuscript. 

\bibliography{Bibliography}

\clearpage
\onecolumngrid

\setcounter{figure}{0}
\setcounter{equation}{0}
\setcounter{section}{0}
\renewcommand\thefigure{S\arabic{figure}}
\renewcommand\theequation{S\arabic{equation}}
\renewcommand\thesection{S-\Roman{section}}
\renewcommand{\theHfigure}{Supplement.\thefigure}
\renewcommand{\theHequation}{Supplement.\theequation}
\renewcommand{\theHsection}{Supplement.\thesection}

\begin{center}
{\Large SUPPLEMENTAL MATERIAL}
\end{center}

\title{Topological Varma superfluid in optical lattice}

\author{M. Di Liberto$^{1,2}$, A. Hemmerich$^{3,4,5}$, and C. Morais Smith$^{1,5}$}
\affiliation{$^1$ Institute for Theoretical Physics, Centre for Extreme Matter and Emergent Phenomena, Utrecht University, Leuvenlaan 4, 3584CE Utrecht, the Netherlands \\
$^2$INO-CNR BEC Center and Dipartimento di Fisica, Universit\`a di Trento, 38123 Povo, Italy\\
$^3$Institut f\"{u}r Laser-Physik, Universit\"{a}t Hamburg, LuruperChaussee 149 22761 Hamburg, Germany \\
$^4$The Hamburg Centre for Ultrafast Imaging, Luruper Chaussee 149, 22761 Hamburg, Germany\\
$^5$Wilczek Quantum Center, Zhejiang University of Technology, Hangzhou, China}

\date{\today}


\maketitle

\section{Quadratic band touching in non-interacting systems}

To understand the findings discussed in the main text, it is instructive to make a comparison with a non-interacting system displaying a quadratic band-touching point (QBTP), for which time-reversal symmetry is broken by adding an appropriate term.

\begin{figure}[bht]
\includegraphics[width=.7\columnwidth]{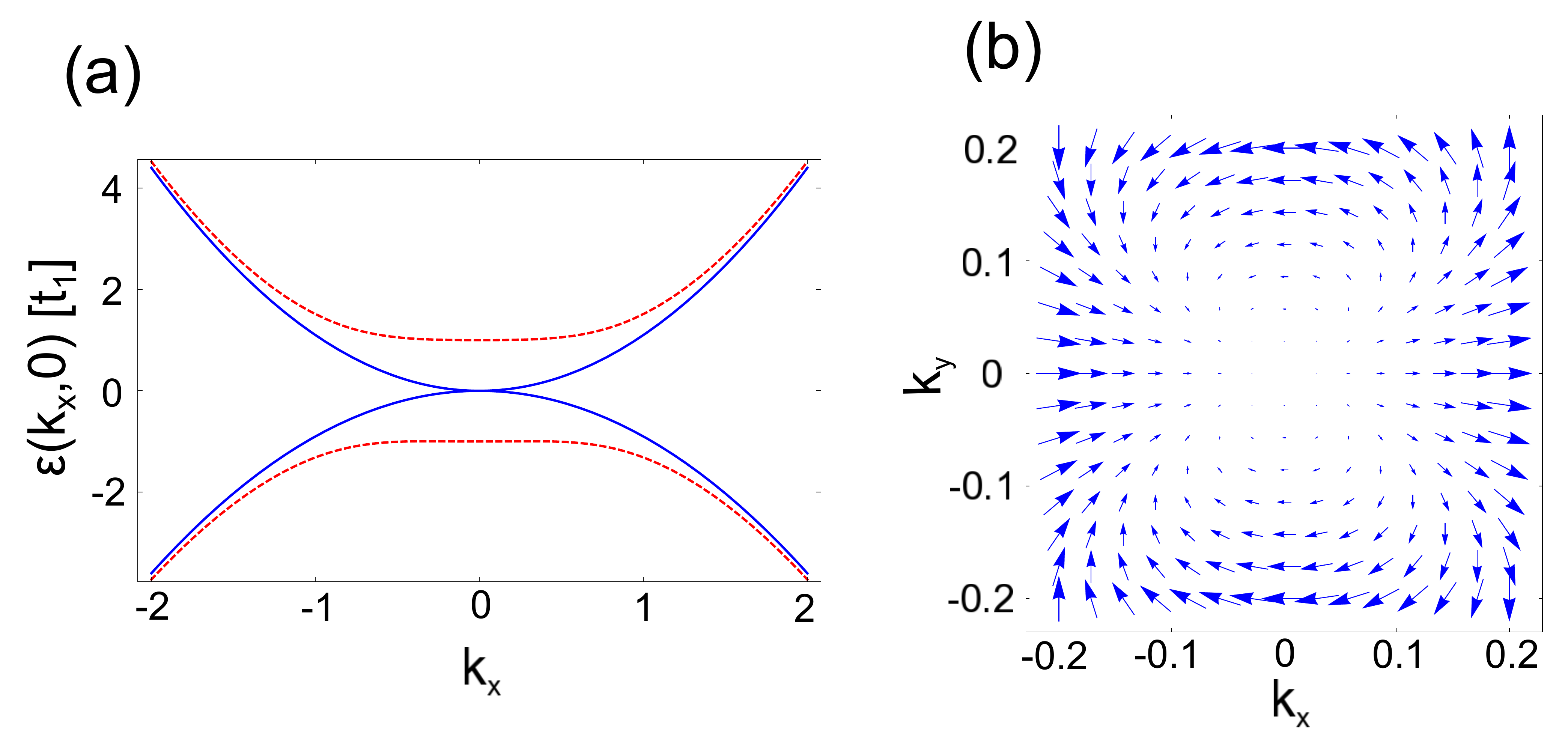}
\caption{(a) Spectrum of the inversion-symmetric two-band model with time-reversal symmetry (continuous line) and without time-reversal symmetry (dashed line) for $k_y = 0$. (b) Vortex structure with winding number $W=2$ of the two-component vector $\vec{h} = (d_1(\b k), d_3(\b k))$. Parameters are $t_1 = 1$, $t_2 = 1$, $t_3 = 2$ and $m=0$ $(m = -1)$ for the time-reversal-symmetry invariant (broken) case.}
\label{QBTP}
\end{figure}

The Hamiltonian describing the low-energy physics of a QBTP can be written in general as
\be
\label{hamqbtp}
H^{0,\rm{eff}}_{\b k} = \left(
\begin{array}{cc}
 k_x k_y t_1+\left(k_x^2+k_y^2\right) t_2 & \left(k_x^2-k_y^2\right)
   t_3 \\
 \left(k_x^2-k_y^2\right) t_3 & -k_x k_y
   t_1 + \left(k_x^2+k_y^2\right) t_2 \\
\end{array}
\right)\,,
\ee
or, using the Pauli matrices and the identity matrix to decompose the 2x2 Hamiltonian,  
\be
H^{0,\rm{eff}}_{\b k} = d_0(\b k)\boldsymbol{1}_{2\times2} + \b d(\b k) \cdot \boldsymbol{\sigma}\,,
\ee
where $d_0(\b k) = (k_x^2+k_y^2)t_2$ and $\b d(\b k) = ((k_x^2-k_y^2)t_3,0,k_x k_y t_1)$. The spectrum of the Hamiltonian is $\epsilon(\b k) = d_0(\b k) \pm d(\b k)$, where $d(\b k) = \sqrt{d_1(\b k)^2 + d_2(\b k)^2 + d_3(\b k)^2}$. In the specific case of the model discussed in the main text, one can perform a unitary transformation of the tight-binding Hamiltonian using as basis the eigenstates at the $M$ point. This allows to decouple the high-energy modes and obtain an effective theory of the form (\ref{hamqbtp}) for the two low-energy modes. The parameters of the effective theory are related to the tight-binding parameters by the relations $(t = 1)$
\bea
t_1 &=& \f{t'}{2} \left( 1 + \f{\xi_p}{\sqrt{16+\xi_p^2}} \right) \,,\\
t_2 &=& t_3 = \f{1}{2 \sqrt{16+\xi_p^2}}\,.\nn
\eea
In the model discussed in the main text, the lowest branch of the dispersion is flat along the directions $k_x = 0$ or $k_y = 0$. This occurs because $t_2 = t_3$, but here we will consider the most general case.

The Hamiltonian is time-reversal and inversion invariant. As discussed in Ref.~\cite{Sun2012}, inversion symmetry $\textrm{U}_I H^{0,\rm{eff}}_{\b k} \textrm{U}_I^{-1} = H^{0,\rm{eff}}_{-\b k}$ requires that the Berry flux $\Phi_B$ relative to the QBTP is defined up to a shift of $4\pi$. Therefore, systems exhibiting a QBTP can belong to two distinct topological classes, namely $\Phi_B = 0$ and $\Phi_B = 2\pi$. The class $\Phi_B = 0$ corresponds to the case where the QBTP occurs as an accidental band crossing, while the class $\Phi_B = 2\pi$ is the topologically non-trivial case. The Hamiltonian in Eq.~(\ref{hamqbtp}) describes the case $\Phi_B = 2\pi$, as one can see from the vortex structure with winding number $W=2$ of the vector $\vec{h} = (d_1(\b k), d_3(\b k))$ in Fig.~\ref{QBTP}(b). The band-degeneracy can be removed by breaking time-reversal symmetry, for instance by adding a non-vanishing $d_2(\b k) = m$. The Berry potential $A_\alpha(\b k) = -i \mv{\psi_{\b k} | \pa_{k_\alpha} | \psi_{\b k}}$ reads

\be
A_\alpha(\b k) = \f{d_1(\b k) \pa_{k_\alpha} d_2(\b k)-d_2(\b k)\pa_{k_\alpha} d_1(\b k)}{2 d(\b k) (d(\b k) \pm d_3(\b k))}\,,
\ee
where $+$ ($-$) denotes the upper (lower) branch of the spectrum. Let us consider, for simplicity, the upper branch. The Berry curvature $F_{xy} = \pa_{k_x} A_y - \pa_{k_y} A_x $ then reads
\be
F_{xy}(\b k) = -\f{m\, t_1 t_3(k_x^2 + k_y^2)}{\left( m^2 + t_1^2 k_x^2 k_y^2 + t_3^2 (k_x^2 - k_y^2)^2 \right)^{3/2}}\,.
\ee
One can notice that the Berry curvature has a general D$_4$ symmetry, except for $t_1 = 2t_3$, for which full rotational symmetry is realised (not shown). For any finite value of the time-reversal-symmetry breaking parameter $m$, the Berry curvature is non-zero for $\b k \neq 0$. It approaches zero at large momenta and vanishes quadratically at $\b k = 0$, whereas it has peaks at finite momenta. This behaviour is different from the one obtained for band structures displaying Dirac cones, where the maximum value of the Berry curvature appears at the gap-opening point. We conclude that the momentum dependence of the dispersion crucially determines the behaviour of the Berry curvature. Moreover, we can also compare the form of the Berry curvature that we found in the interacting system (Fig.~4(c) in the main text) with the results displayed here (Fig.~\ref{QBTP}(a)) and observe that the form of the Berry curvature is analogous: it is characteristic of a system with a QBTP where the gap has opened as a consequence of a time-reversal symmetry-breaking mechanism.

\begin{figure}[!t]
\label{BCurv_eff}
\includegraphics[width=.7\columnwidth]{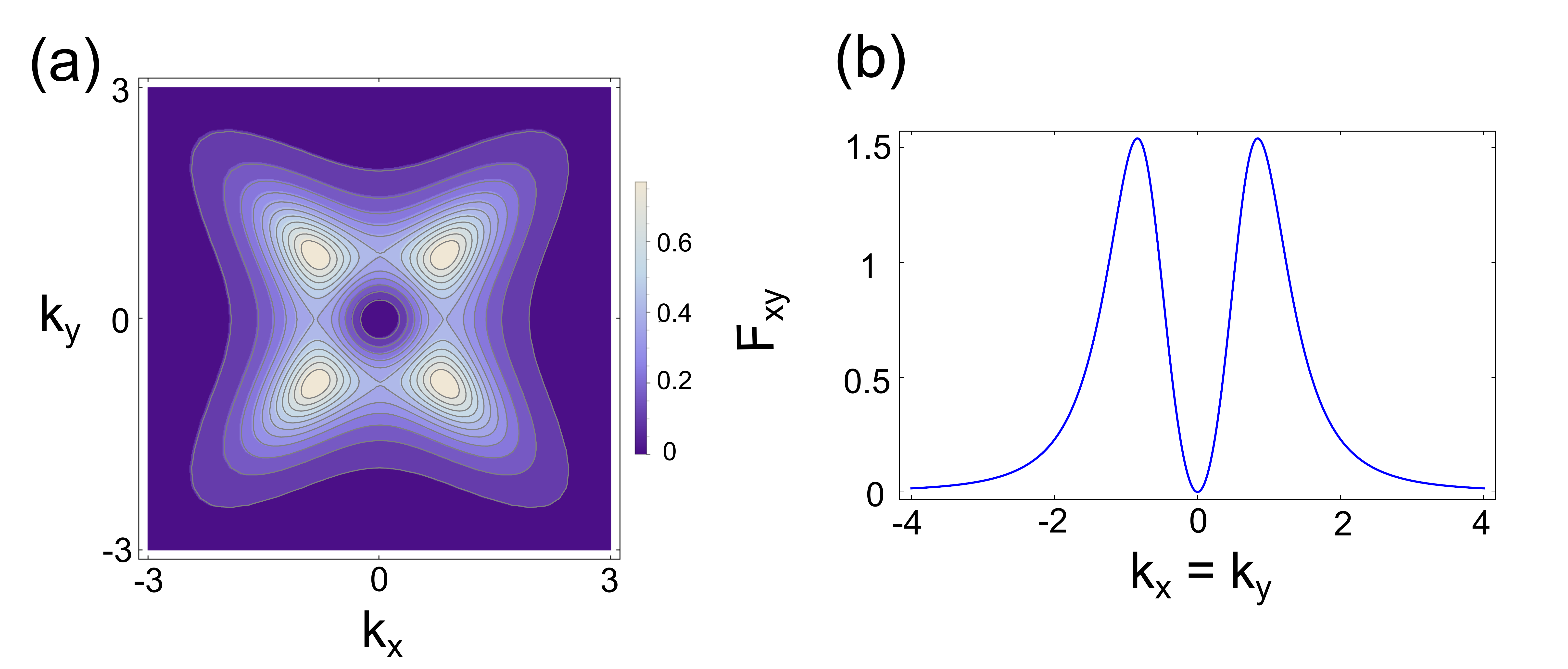}
\caption{ (a) Berry curvature of the time-reversal symmetry broken two-band model. (b) Cut of the Berry curvature for $k_x = k_y$. Parameters are chosen as in Fig.~\ref{QBTP}.}
\end{figure}

 
{\section{Mean-field theory.}}  


Here, we present more details of the mean-field calculations describing the ground-state physics in the main text. The form of the eigenstates at the minima of the spectrum allows us to take an ansatz for the condensate wave function of the form
\begin{align}
\mv{p_{\alpha,\b{r}}} =& \sqrt{\rho_p} e^{i\b{k}_{\alpha}\cdot \b{r}} e^{i\theta_\alpha} \equiv \sqrt{\rho_p} e^{i\theta_{\alpha,\b{r}}} \\
\mv{s_{\alpha,\b{r}}} =& \sqrt{\rho_s} e^{i\b{k}_\alpha\b{\cdot (r}+\b{e}_\alpha)} e^{i\phi_\alpha} \equiv \sqrt{\rho_s} e^{i\phi_{\alpha ,\b{r}}} \,,
\end{align}
where we let the condensate density on $p$ and $s$ orbitals be different in general, and we introduced global phases to be fixed by minimizing the mean-field energy. For the condensate at the $M$ point, one trivially has that $\b k_x=\b k_y = (\pi,\pi)$, whereas for the condensate at $X$ and $X'$, the ansatz requires $\b k_x= (\pi,0)$ and $\b k_y = (0,\pi)$. This ansatz is justified by the fact that the eigenstate at the $X$ point has non-vanishing components only for $p_x$ and $s_x$ orbitals, while the eigenstate at the $X'$ point analogously has non-zero components only for $p_y$ and $s_y$ orbitals. 

The angular momentum, in the mean-field approximation, takes the simple form
\be
\label{angm}
\mv{L_{z,\b r}} = 2 \rho_p \sin(\theta_{y,\b{r}} - \theta_{x,\b{r}})\,.
\ee
Therefore, the mean-field energy contribution coming from the interaction is
\be
E^{\rm{MF}}_{\rm{int}} \simeq  \f{\rho^2_pU_p}{3} \left\{  5+  \cos[2\Delta\theta_{\b{r}}]\right\} + U_s \rho_s^2\,,
\ee
where we defined $\Delta\theta_{\b{r}} = \theta_{x,\b{r}}-\theta_{y,\b{r}}$. One immediately notices that interactions are minimized when $\Delta\theta_{\b{r}} = \pm \pi/2$, which in turn implies that the expectation value of the angular momentum operator $\mv{L_{z,\b r}}$ is non-zero, as a consequence of the time-reversal-symmetry breaking. 

Let us now define the condensation points as $\b{k}_\alpha \equiv (k_{\alpha, x}, k_{\alpha, y})$. The mean-field energy contribution coming from the hopping terms reads
\begin{align}
E^{\rm{MF}}_{\rm{kin}} = 4t \sin(k_{x,x}/2) \sqrt{\rho_s\rho_p} \sin(\phi_x-\theta_x) + 4t \sin(k_{y,y}/2) \sqrt{\rho_s\rho_p} \sin(\phi_y-\theta_y)\nn\,.
\end{align}
Notice that there is no contribution from the hopping term $t'$. This term will, however, be present in the study of the excitations. Moreover, for the two kinds of condensate considered here, we have $\sin(k_{x,x}/2) = \sin(k_{y,y}/2) = 1$.

Assuming $t>0$, the minimization of the kinetic energy requires
\be
\phi_\alpha-\theta_\alpha = -\pi/2\,.
\ee
In the rest of the section we will adopt the gauge choice $\phi_y = 0$.\\


\subsection{Condensate in $X = (\pi,0) $ and $X'=(0,\pi)$}

In this case, since $\b k_x=(\pi,0)$ and $\b k_y = (0,\pi)$, one has $2\left(\theta_{x,\b{r}}-\theta_{y,\b{r}}\right) =2 (\b k_x\cdot\b r- \b k_y \cdot\b r + \theta_x-\theta_y )= 2\pi(m-n) + 2 (\theta_x-\theta_y) = \pm\pi$, where we defined $\b r \equiv (m,n)$, with $m,n$ integers. Since the shift $2\pi(m-n)$ does not change the value of the interaction term (proportional to $\cos(2\Delta\theta_{\b r})$), we find the solution for the global phases
\be
\theta_x = \pi\,,\quad \theta_y = \pi/2\,,\quad \phi_x = \pi/2\,,\quad \phi_y = 0\,.
\ee
The mean-field free energy is 
\be
E^{\rm{MF}} \simeq -8t \sqrt{\rho_s\rho_p}+\f{4}{3}\rho^2_pU_p + U_s \rho_s^2+2\xi_p\rho_p\,,
\ee 
and the order parameters can be written as 
\begin{align}
\mv{p_{x,\b{r}}} =&- \sqrt{\rho_p} e^{i\b k_x\cdot \b{r}}\,, \nn\\
\mv{p_{y,\b{r}}} =& \, i\sqrt{\rho_p} e^{i\b k_y\cdot \b{r}}\,, \nn\\
\mv{s_{x,\b{r}}} =& - \sqrt{\rho_s} e^{i\b k_x\cdot \b{r}} \,, \nn\\
\mv{s_{y,\b{r}}} =& \, i \sqrt{\rho_s} e^{i\b k_y\cdot \b{r}}\,.
\end{align}
The angular momenta on the $\cal P$ sites are non-zero and have an anti-ferromagnetic structure,
\be
\mv{L_{z,\b r}} = -2\rho_p \sin[\pi(m-n) +\pi/2] = -2\rho_p (-1)^{m+n}\,.
\ee

In this ground state, because of the staggered structure, translational symmetry is broken. Let us now check the pattern of currents. We introduce a bond-current operator $J^{\mu\nu}(i,j)$ that describes the superfluid current between site $i$ (flavor $\mu$) and site $j$ (flavor $\nu$)
\be
\label{current}
J^{\mu\nu}(i,j)=-i t_{ij}(b^\dag_{\mu, i}b_{\nu, j} - b^\dag_{\nu, j} b_{\mu, i}).
\ee
The expectation value of this operator appears, for instance, in the continuity equation obtained from the Gross-Pitaevskii equation on the lattice. We will therefore calculate the average superfluid current of each bond using the mean-field ansatz $b_{\nu, j} \rightarrow \mv{b_{\nu, j} }$. 

The currents along the edges vanish, whereas the currents along the diagonal $s-s$ bonds are finite and read
\begin{align}
\mv{J^{s_xs_y}(\b r& + \b e_x,\b r + \b e_y)} = i t' \rho_s(-i - i ) (-1)^{m+n}= 2t'\rho_s (-1)^{m+n}\nn\,\\
\mv{J^{s_xs_y}(\b r& + \b e_x,\b r + 2\b e_x + \b e_y)} = i t' \rho_s(i + i ) (-1)^{m+n}= -2t'\rho_s (-1)^{m+n}\,,  \nn\\
\mv{J^{s_ys_x}(\b r& + \b e_y,\b r + 2\b e_y + \b e_x)} = i t' \rho_s(-i - i ) (-1)^{m+n}= 2t'\rho_s (-1)^{m+n}\,,  \nn\\
\mv{J^{s_ys_x}(\b r& + 2 \b e_x + \b e_y,\b r + 2\b e_y + \b e_x)} = i t' \rho_s(i + i )(-1)^{m+n} = -2t'\rho_s(-1)^{m+n}\,.
\end{align}
The current pattern is shown in Fig.~3(a) of the main text. In the enlarged unit cell, one can see that inversion symmetry (with respect to any $\cal P$ site) is not broken, but mirror symmetry and translation are. 

\subsection{Condensate in $M = (\pi,\pi) $}

In this case, since $\b k_x=\b k_y = (\pi,\pi)$, one has $\theta_{x,\b{r}}-\theta_{y,\b{r}} = \theta_x-\theta_y = \pm \pi/2$. Let us consider the solution with the plus sign because the other solution is its time reversal conjugate. We obtain for the phases, as before,
\be
\theta_x = \pi\,,\quad \theta_y = \pi/2\,,\quad \phi_x = \pi/2\,,\quad \phi_y = 0\,.
\ee
The mean-field free energy is 
\be
\label{mfenMp}
E^{\rm{MF}} \simeq -8t \sqrt{\rho_s\rho_p}+ \f{4}{3}\rho^2_pU_p + U_s \rho_s^2 + 2 \xi_p \rho_p\,,
\ee
and the order parameters can be written as 
\begin{align}
\mv{p_{x,\b{r}}} =& - \sqrt{\rho_p} e^{i\bm{\pi}\cdot \b{r}}\,, \nn\\
\mv{p_{y,\b{r}}} =& \,i\sqrt{\rho_p} e^{i\bm{\pi}\cdot \b{r}}\,, \nn\\
\mv{s_{x,\b{r}}} =& - \sqrt{\rho_s} e^{i\bm{\pi}\cdot \b{r}} \,, \nn\\
\mv{s_{y,\b{r}}} =& \, i\sqrt{\rho_s} e^{i\bm{\pi}\cdot \b{r}}\,.
\end{align}

The angular momenta on the $\cal P$ sites are non-zero and have a ferromagnetic structure
\be
\mv{L_{z,\b r}} = -2\rho_p\,.
\ee
Let us now calculate the bond currents. We start from the $p_x-s_x$ bond, 
\be
\mv{J^{p_xs_x}(\b r, \b r + \b e_x)} = it \sqrt{\rho_s \rho_p} \left(1 - 1\right) = 0\,.
\ee
The same holds for the bonds $p_y-s_y$, so there are no currents running along the edges of the squares. The currents can only be among the $s-s$ bonds, and there are four of these bonds in each plaquette,
\be
\mv{J^{s_xs_y}(\b r + \b e_x,\b r + \b e_y)} = i t' \rho_s(-i - i ) = 2t'\rho_s \,.
\ee
This is the current that goes from $\b r + \b e_x$ to $\b r+ \b e_y$, the direction of which depends on the sign of $t'$. The other currents in the plaquette read
\begin{align}
\mv{J^{s_xs_y}(\b r + \b e_x,\b r + 2\b e_x + \b e_y)} &= i t' \rho_s(i + i ) = -2t'\rho_s\,,  \nn\\
\mv{J^{s_ys_x}(\b r + \b e_y,\b r + 2\b e_y + \b e_x)} &= i t' \rho_s(-i - i ) = 2t'\rho_s\,,  \nn\\
\mv{J^{s_ys_x}(\b r + 2 \b e_x + \b e_y,\b r + 2\b e_y + \b e_x)} &= i t' \rho_s(i + i ) = -2t'\rho_s\,.
\end{align}

In Fig.~3(b) of the main text, the pattern of superfluid currents and angular momenta (represented by a minus sign) is shown. This ground state breaks time-reversal and mirror symmetry but does not break translational and inversion symmetry. Moreover, in the region of the plaquette delimited by the bond currents, there is a net flux piercing the surface, while the total flux per plaquette is zero (see Fig.~3(b) in the main text). The global phase in $\b r + \b e_x$ is $\pi$, whereas in $\b r + \b e_y$ it is $\pi/2$. Therefore, the phase difference across each bond is $\Delta\phi_{\rm{bond}} = \pi/2$ and the total phase picked up in a closed path (corresponding to the flux) is $\Phi = 2\pi$. The main difference between this phase and the phase corresponding to condensation in $X-X'$ is the fact that translational symmetry is preserved here.


 \begin{figure}[t]	
 	\begin{center}
 	\includegraphics[width=0.7\columnwidth]{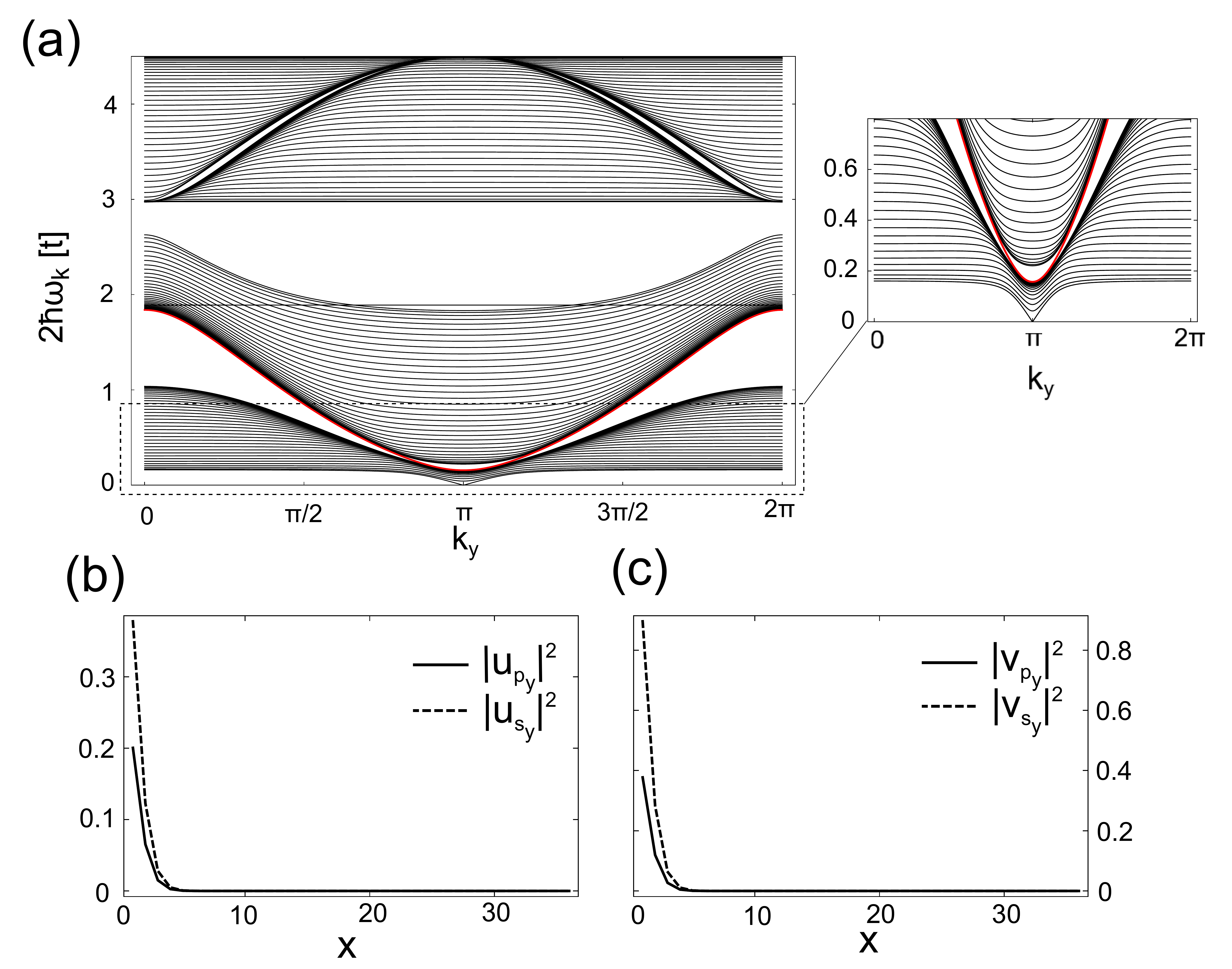}
 	\end{center}
\caption{(a) Bogolybov spectrum on a cylindrical geometry with $L=35$ unit cells, in units where $t=1$, $t'=0.2$, $\xi_p=1.5$ and $U=0.005$. The edges have been chosen by cutting along the lattice links connecting the $p-s_y$ orbitals. The red line corresponds to the edge-state dispersion. (b)-(c) Square modulus of the particle ($u(x)$) and hole ($v(x)$) components of the edge-state wavefunction for $k_y = 2.62$. One can clearly observe the localization of the excitation at one edge of the cylinder, with coordinates $x=0$. Analogously, the other branch of the edge-state dispersion for $k_y>\pi$ yields states localized at $x = L$ (not shown). The components of the wave function for $p_x$ and $s_x$ are vanishingly small, and for this reason they are not plotted.}
 	\label{bogspec}
 \end{figure}

{\section{Interactions in real space.}} 


To calculate the excitation spectrum within Bogolyubov theory, each bosonic operator is split into its mean-field expectation value and the fluctuations around it, namely $\phi_{\b r} \rightarrow \mv{\phi_{\b r}} + \delta \phi_{\b r}$. Only terms up to quadratic order in the fluctuations are retained. Imposing that the terms linear in the fluctuations vanish yields a constraint for the chemical potential. 

The quadratic terms coming from the interactions on the $\cal S$ sites are
\begin{align}
\f{U_s}{2} \sum_{\b r\in \cal S}\left[ \left(\mv{s_{\b r}}^2\delta s_{\b r}^\dagger \delta s_{\b r}^\dagger + \textrm{h.c.} \right)+ 4 |\mv{s_{\b r}}|^2 \delta s_{\b r}^\dagger \delta s_{\b r}\right]\,.
\end{align}
On the $\cal P$ sites there are more possibilities and one finds
\begin{align}
&\f{U_p}{2} \sum_{\b r\in \cal P} \left\{\left[ \left( \mv{p_{x,\b r}^\dagger}^2 + \f1 3 \mv{p_{y,\b r}^\dagger}^2 \right)\delta p_{x,\b r} \delta p_{x,\b r} +  \left( \mv{p_{y,\b r}^\dagger}^2 + \f1 3 \mv{p_{x,\b r}^\dagger}^2 \right) \delta p_{y,\b r} \delta p_{y,\b r}   \right.\right. \nn\\
&\left.+ \f 4 3 \mv{p_{x,\b r}^\dagger} \mv{p_{y,\b r}^\dagger} \delta p_{x,\b r}\delta p_{y,\b r} + \f 4 3 \left(  \mv{p_{x,\b r}^\dagger} \mv{p_{y,\b r}} +  \mv{p_{y,\b r}^\dagger} \mv{p_{x,\b r}} \right) \delta p_{x,\b r} \delta p_{y,\b r}^\dagger \right] + \rm{h.c.} \nn\\
&+\left(  4\mv{p_{x,\b r}^\dagger} \mv{p_{x,\b r}} +  \f 4 3 \mv{p_{y,\b r}^\dagger} \mv{p_{y,\b r}} \right) \delta p_{x,\b r}^\dagger \delta p_{x,\b r}  \nn\\
& \left. + \left(  4\mv{p_{y,\b r}^\dagger} \mv{p_{y,\b r}}+  \f 4 3 \mv{p_{x,\b r}^\dagger} \mv{p_{x,\b r}} \right) \delta p_{y,\b r}^\dagger \delta p_{y,\b r} \right\} \,.
\end{align}
The Hamiltonian has therefore become quadratic. On can now plug in the values of the order parameters and Fourier transform to obtain the Bogolyubov Hamiltonian shown in the main text. The results in a cylindrical geometry are shown in Fig.~\ref{bogspec}(a). In addition, the square modulus of the components of the edge-state wavefunction are shown in Figs.~\ref{bogspec}(b) and (c). 
 \begin{figure}[!ht]	
 	\begin{center}
 	\includegraphics[width=0.4\columnwidth]{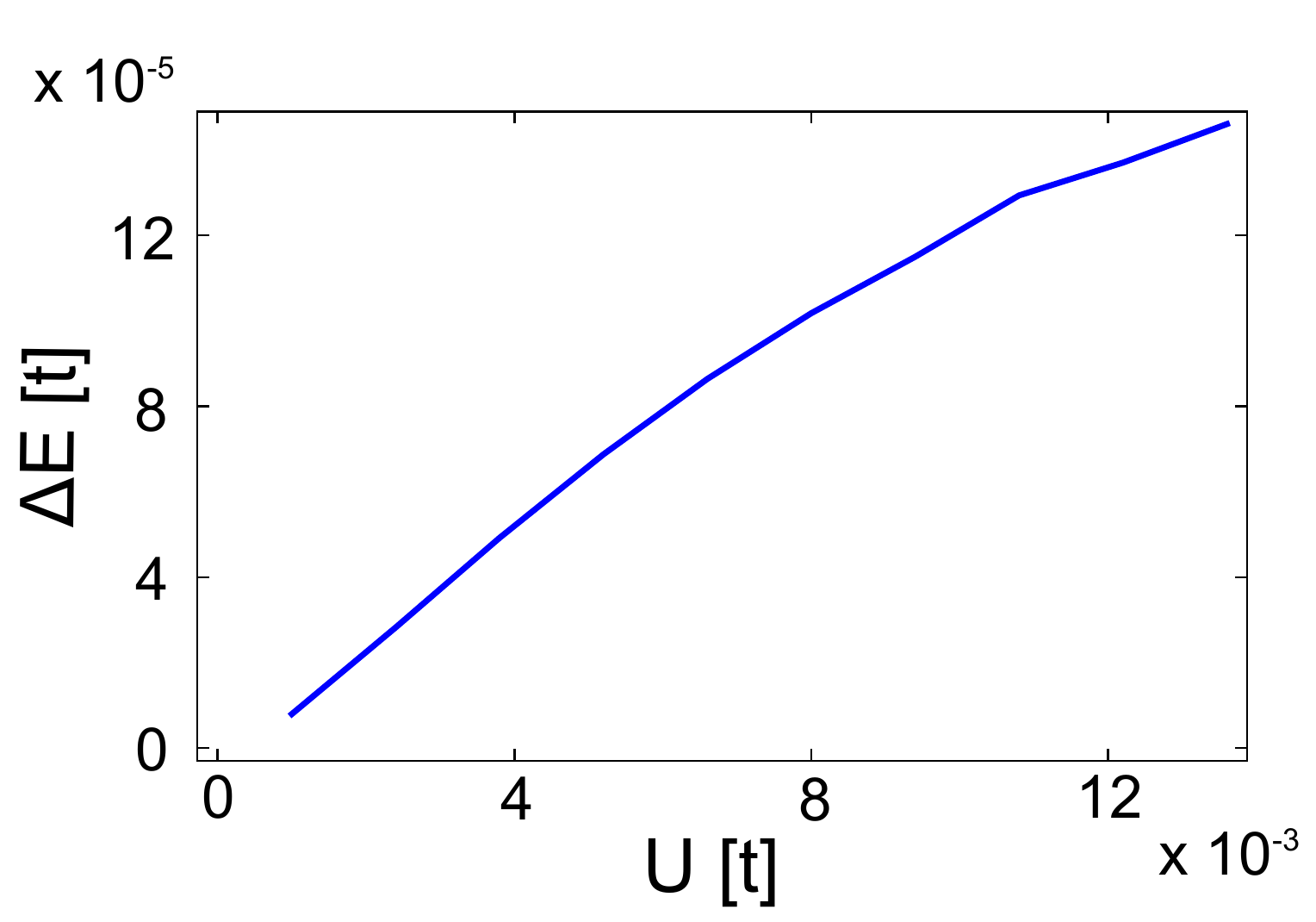}
 	\end{center}
 	\caption{Quantum corrections to the ground state energy. Parameters are chosen in units where $t=1$, $t'=0.2$, and $\xi_p = 1.5$.  }
 	\label{Egs}
 \end{figure}

{\section{Quantum fluctuations.}} 


The degeneracy of the two ground states discussed in the main text can be removed by including the corrections to the mean-field energy arising from the zero-point fluctuations given by the phonon modes (see Ref.~\cite{Zhai2012} for a description of the procedure), 
\be
\Delta E = E^{M} - E^{X-X'} = \f 1 2 \sum_i \int \f{d\b k}{4\pi^2} \, \Delta\hbar\omega^{i}_{\b k}\,.
\ee
In Fig.~\ref{Egs}, it is shown that the quantum corrections favour the $X-X'$ condensate, which is therefore the lowest-energy state. However, the energy difference is very small (of the order of 1pK). We expect that in each experimental run the metastable $M$-condensate will be therefore randomly populated and will decay to the true vacuum for sufficiently long times. The life-time of the $M$ condensate will depend on the details of the interactions, namely density and phase fluctuations. The estimation of this quantity is beyond the scope of this work. \\


{\section{Ground-state conductivity.}} 


To investigate the possibility that the ground states discussed in the main text possess non-trivial topological properties, we followed the standard procedure of twisting the boundary conditions of the lattice by inserting fluxes $\b{\Theta}=(\Theta_x,\Theta_y)$ \cite{Avron1985,Lind2011} to calculate the Hall response. The transverse (Hall) conductivity is then given by $\sigma_{xy}\propto \int_0^{2\pi}\int_0^{2\pi}d^2\b{\Theta}\, \rm{Im}\mv{\pa_{\Theta_x}\Psi |\pa_{\Theta_y}\Psi }$, where $|\Psi\ket$ is the wave-function of the many-body ground state. 

We performed imaginary-time Gross-Pitaevskii calculations on small lattices to determine the ground-state wave function in the presence of twisted boundary conditions. We assumed a mean-field product state $|\Psi\ket \equiv \prod_i |\psi_i\ket$, where $|\psi_i\ket$ are on-site wave functions represented as coherent states. We applied the method presented in Ref.~\cite{Hatsu2005} to calculate the integral in flux space and we found zero conductivity. 

To interpret this result, we notice that the ground-state shown in Fig.~3 can be seen as a vortex lattice in a superfluid. In the presence of a single vortex, it has been shown that a finite non-trivial Hall conductivity will appear in a bosonic lattice system. This phenomenon is a consequence of the fact that the vortex will flow up-stream or down-stream when a bias current is applied, thus causing a phase gradient orthogonal to its motion which ultimately leads to a Hall response of the superfluid \cite{Lind2011}. Here, in the presence of a vortex lattice, the conductivity averages to zero because the contribution of all the vortices cancels out. The $X-X'$ and $M$ ground states are therefore topologically trivial.

The presence (or absence) of Hall response of the ground state is conceptually different from the Hall response of the Bogolyubov excitations. The bogolons are quasi-particles behaving as free bosons. Because of the non-vanishing Chern number of the Bogolyubov band, and therefore the non-vanishing Berry curvature, the semi-classical dynamics of the bogolons in the presence of a drag force is affected by the presence of an anomalous velocity, which is at the origin of the Hall current \cite{Niu2010}.


\end{document}